\documentstyle[11pt,epsfig]{article}
\newcommand{\beq}{\begin{equation}}
\newcommand{\eeq}{\end{equation}}
\newcommand{\beqa}{\begin{eqnarray}}
\newcommand{\eeqa}{\end{eqnarray}}

\date{}
\title{Fragment Formation and Phase Transitions in Heavy Ion Collisions}
\author{{\normalsize T. Gaitanos, H. H. Wolter (Sektion Physik, Universit\"at 
M\"unchen, Germany),}\\
{\normalsize C. Fuchs (Universit\"at T\"ubingen, Germany)}}

\begin{document}
\baselineskip15pt

\maketitle

\begin{abstract}
We investigate the thermodynamical properties of nuclear matter in heavy ion 
collisions, in particular, with respect to questions of thermodynamical instability, phase transitions and fragmentation. For this we analyze results of relativistic transport calculations of $Au+Au$ collisons at intermediate energies, separately for spectator and participant matter. On one hand, we determine local thermodynamical variables from the analysis of the local momentum distribution; on the other, we analyze fragment energy spectra in a blast model scenario. We find that the spectator represents an instable, equilibrized fragmenting source, while in the participant no such common source can be identified. Our results compare well with experimental determinations of temperatures and flow velocities.
\end{abstract}

\subsection*{Introduction}

Heavy ion collisions in the energy range up to a few GeV per nucleon have been investigated to obtain information on the nuclear matter equation-of-state (EOS) in a domain where hadronic degrees of freedom are thought to be the appropriate description. In such collisions several aspects of the EOS can be explored: high density and temperature in the compression phase and low density and possible signals of liquid-gas phase transitions in the expansion and the final fragmentation phase. Much work has been performed on the first question both experimentally, see e.g.\cite{reisdorf}, and theoretically, e.g. \cite{blaett93}, by investigating in particular flow observables but also particle production \cite {cassing}. The investigation is complicated by the fact that at this stage equilibrium is not generally reached in medium energy collisions and therefore non-equilibrium effects have to taken into account in extracting the high density EOS \cite{gait99}.

In this contribution we discuss  the information that can be obtained in the expansion phase. Since the NN-interaction is qualitatively similar to inter-molecular forces with short range repulsion and longer range attraction, a phase diagram similar to a van-der-Waals system is to be expected. Thus there should exist first order phase transitions in bulk nuclear matter; however, the question is how such phase transitions occur in finite systems and what signals can be observed \cite{chomaz}. The expanding system will finally have to decay into fragments of different sizes (called multi-fragmentation) and information is sought from experimental measurements of fragment distributions, spectra, correlations, etc. In the interpretation of fragment spectra often the concept of an equilibrated freeze-out configuration is used, which is the point in the evolution of the system, where the fragments are formed and essentially cease to interact (except for Coulomb interactions). An important question is whether such a configuration exists, whether it is in thermal equilibrium, characterized by a unique temperature, and in chemical equilibrium, such that all fragment species are formed at the same instant. 

In this work the approach to answer these questions is to perform transport calculations evolving the system microscopically from the initial stage through the formation of fragments. Such calculations are based on non-equilibrium formulations \cite{malfl} and no concept of equilibrium, freeze-out, etc. is neccessary. However, one may analyze the calculation with respect to questions, like the degree of equilibrium, whether a temperature can be defined, whether the system enters a region of instability, etc. One can also analyze the final fragmenting state using the same procedures as in experiment. By comparing the results of these two approaches we hope to understand better the questions involved in phase transitions in the finite system of heavy ion collisions.

Transport calculations are usually done in a semiclassical approximation on the mean field level. However, fluctuations and correlations beyond the mean field level are important in the thermodynamically instable situation of a phase transition. In this work the approach is used \cite{colonna} to gauge the numerical fluctuations arising from the use of the test particle method in such a way as to simulate the physical fluctuations. To go beyond this approach fluctuations have been included in the Boltzmann-Langevin approach \cite{randrup}, but mainly schematic calulations have been done with this method. Recently together with the Catania group we have proposed a new and practical way to include fluctuations in a physical way in transport calculations \cite{fluct}. This procedure is not used here but will be employed in the future in somewhat less empirical investigation of fragmentation phenomena. 

\subsection*{Thermodynamical analysis}

We base our investigation on relativistic transport calculations of the 
Boltz\-mann-Nord\-heim-Vlasov type \cite{blaett93,gait99}. The equation of motion for the one-body distribution function $f(x,p^* )$ reads
\beqa
&&  \left[p^{*\mu} \partial_{\mu}^x  + \left( p^{*}_{\nu} F^{\mu\nu}     
+ m^* \partial^{\mu}_x m^* \right) 
\partial^{p^*}_{\mu} \right] f(x,p^* ) = \rm{I}_{coll}
\quad .
\label{TP5}
\eeqa
with the kinetic momentum $p^{*\mu} = p^{\mu}+\Sigma^{\mu}$ and the effective mass $m^* = m + \Sigma_s$, where $\Sigma^{\mu}$ and $\Sigma_s$ are the scalar and vector self energies, respectively.  $\rm{I}_{coll}$ is a Boltzmann-type collision term respecting the Pauli-principle. For the self 
energies  we have adopted the non-linear parametrization NL2 \cite{blaett93}. In ref. \cite{gait99} we compared this 
parametrization to more realistic non-equilibrium self energies based on 
Dirac-Brueckner calculations. With respect to the thermodynamical variables 
discussed here, we found no essential differences between the two models, 
and thus we use the simpler NL2 here. The calculations are performed in the relativistic Landau-Vlasov approach which was described in 
detail in ref. \cite{fuchs95}. It uses Gaussian test particles in coordinate and 
momentum space and thus allows to construct locally a smooth momentum distribution.

The local thermodynamical properties are obtained from an analysis of the momentum dependence of $f(x,p^* )$ at a given space-time point $x$. The local 4-current $j^{\mu}$ and the local invariant density $\rho_0=\sqrt{j_\mu j^\mu}$ is directly calculated from  $f(x,p^* )$, furthermore the energy density $\epsilon$ and the pressure $P$ are obtained from the energy momentum tensor. 
For the determination of a local temperature the local momentum distribution is subjected to a fit in terms of covariant 
hot Fermi--Dirac distributions of the form  \cite{fuchs97}
\begin{equation}
n(x,{\vec p},T) = 
\frac{1}{ 1+ \exp\left[ - (\mu^* - p_{\mu}^* u^\mu )/T \right]}
\label{fermi}
\end{equation}
with the temperature $T$ and the effective chemical potential $\mu^* (T)$ .  
The local streaming 
four-velocity $u_\mu$ is given as 
$u^{\mu} = j^{\mu}/\rho_0$. Then 
the temperature $T$ is the only fit parameter to be directly 
determined from the phase space distribution. This temperature is a local thermodynamic 
temperature, which in the following we denote as $T_{loc}$. 
Expression (\ref{fermi}) is appropriate for a system in local 
equilibrium. We model the anisotropic momentum distributions in the earlier stages of a heavy ion collision  by counter-streaming 
or 'colliding' nuclear matter \cite{fuchs97,sehn96,tuebingen95}, 
i.e., by a superposition of two Fermi distributions   
$ n^{(12)} = n^{(1)} + n^{(2)} -\delta n^{(12)}$,
where $\delta n^{(12)} = \sqrt{n^{(1)} \cdot n^{(2)} }$ guarantees the Pauli principle 
and provides a smooth transition to one equilibrated system. 
In \cite{fuchs97,gait99} it has been 
demonstrated that this ansatz 
allows a reliable description of the participant and spectator 
matter at each stage of the reaction. 

An example of such an analysis is shown in Fig. 1 for a semi-central $Au+Au$ collision. The local momentum distribution is given for characteristic space-time points in the fireball (participant) and in the spectator region. Also the fit is shown that determines the local temperature $T_{loc}$. The participant region is not in equilibrium at the early times shown, however, the spectator represents a well equilibrated system, as does also the participant at later times.
\begin{figure}
\begin{center}
\setlength{\unitlength}{1.0cm}
\begin{picture}(14,8)
\put(1.3,-0.5){\makebox{\epsfig{file=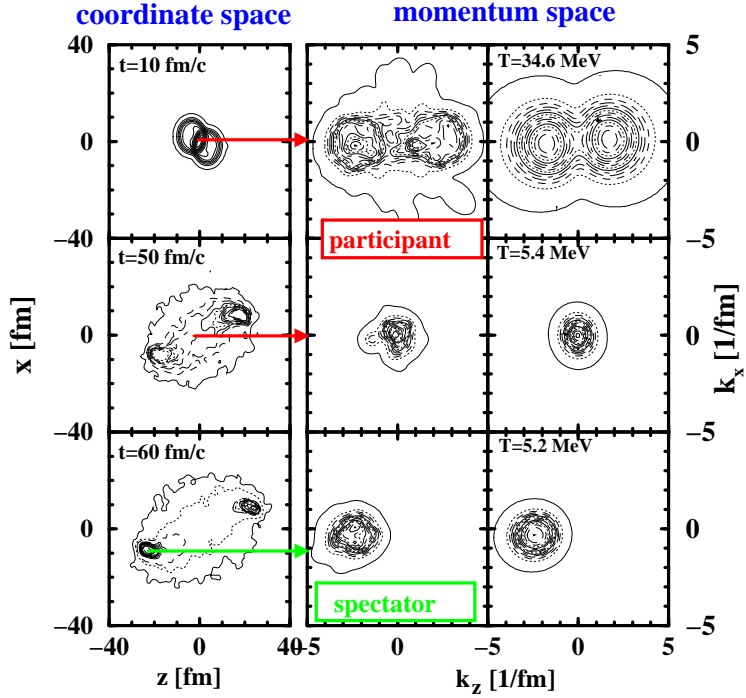,width=10.0cm}}}
\end{picture}
\end{center}
\caption{ Phase space evolution of a semi--central $Au+Au$ collision at 600 AMeV and b=4.5 fm. The left column shows the density distributions at certain times, the two right columns the momentum distribution at the center of mass (upper and middle row) and in the center of the spectator (lower row). The rightmost column shows the fit to determine the local temperature.}
\label{fig1}
\end{figure}

Experimentally much of the information about the thermodynamical behaviour in 
heavy ion collisions originates from the analysis of fragment observables. The physical and practical procedure how to properly decribe fragment 
production is still very much debated \cite{colonna,fluct}. Here we 
 use the simplest algorithm, namely a 
coalescence model, as we have decribed it in ref. 
\cite{gait99}. In brief, we apply phase space coalescence, i.e. nucleons form a 
fragment, if their positions and momenta ($\vec{x}_{i},\vec{p}_{i}$) satisfy
$| \vec{x}_{i}-\vec{X}_f | \leq R_c$ and  
$| \vec{p}_{i}-\vec{P}_f |  \leq P_c$. $R_c,P_c$ are parameters which are 
fitted to reproduce the observed mass distributions and thus guarantee 
a good overall description of the fragment multiplicities. 

Fragment kinetic energy spectra have been analyzed experimentally 
in the Siemens-Rass\-mus\-sen or blast 
model \cite{eos95,reisdorf,siemens}. In this model the kinetic energies are 
interpreted in terms of a thermalized freeze-out configuration, characterized by 
a common temperature and a radial flow, i.e. by an isotropically expanding 
source. In this model the kinetic energies are given by 
\begin{eqnarray}
\frac{dN}{dE} \sim pE \: \int \beta^2 d\beta n(\beta) \exp(\gamma E/T)
 \: \Big[ \frac{ \sinh{\alpha} }{ \alpha }\left( \gamma+\frac{T}{E} 
\right) - 
\frac{T}{E} \cosh{\alpha} \Big],
\label{fit}
\qquad 
\end{eqnarray}
where $p$ and $E$ are the center of mass momentum and the total energy of the 
particle with mass $m$, respectively, and  where $\alpha=\gamma \beta p/T$. 
The flow profile $n(\beta)$ is usually well 
parametrized as a Fermi-type function \cite{reisdorf}; however, the 
results are not very different when using a 
single flow velocity, i.e. $n(\beta) \sim \delta(\beta-\beta_f)$.  One then has two parameters in the fit, namely 
$\beta_f$ and the temperature parameter in eq.(\ref{fit}), which we call 
$T_{slope}$. It is, of course, not obvious that $T_{slope}$ represents a 
thermodynamical 
temperature, in fact, here we aim to find its 
significance. The expression (\ref{fit}) has been applied to kinetic energy 
spectra of all fragment masses simultaneously, yielding a global 
$T_{slope}(global)$, 
or to each fragment mass separately, giving $T_{slope}(A_f)$. Examples of fits to the generated fragment spectra are shown in Fig. 2 for a central $Au+Au$ collision. Spectra are shown with the coalescence model and with the additional condition of a positive binding energy. It is found that the temperature $T_{slope}$ extracted is almost independent of this (however, the number of fragments decreases due to sequential decay).
\begin{figure}
\begin{center}
\setlength{\unitlength}{1.0cm}
\begin{picture}(14,6)
\put(0.5,-0.5){\makebox{\epsfig{file=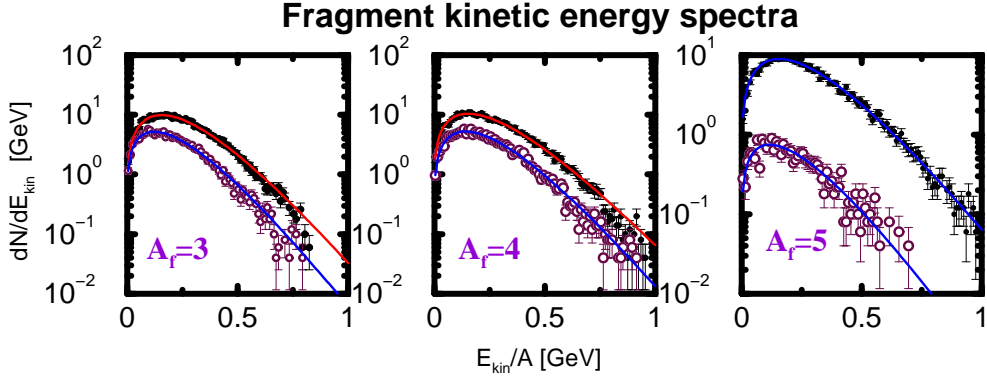,width=12.0cm}}}
\end{picture}
\end{center}
\caption{Fragment kinetic energy spectra for a central $Au+Au$ collisions at 600 AMeV for several fragment masses $A_f$. These are shown for the coalescence model (full points) and with an additional binding energy criterion (open points). The lines represent the blast model fits to determine the slope temperature.}
\label{fig2}
\end{figure}

\subsection*{Spectator matter}

The spectator is that part of the system which has not collided with the other 
nucleus, but which is nevertheless excited due to the shearing--off of part of 
the nucleus and due to absorption of participant particles. We expect from Fig. 1  that it represents a well 
equilibrated piece of nuclear matter at finite temperature.  
\begin{figure}[ht]
\unitlength1cm
\begin{center}
\begin{picture}(14,7.5)
\put(0.5,-0.5){\makebox{\epsfig{file=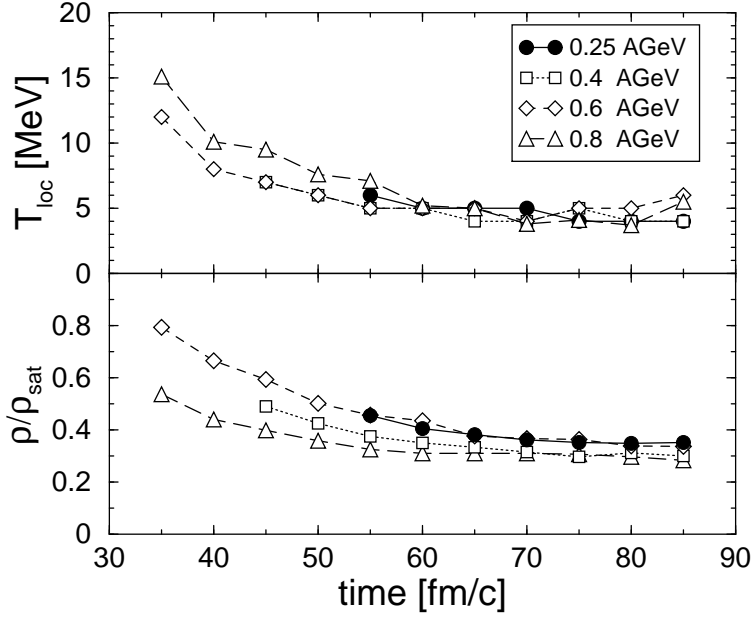,width=10.0cm}}}
\end{picture}
\end{center}
\caption{ Local temperature (top) and density (bottom) evolution in the 
spectator in semi--central $Au+Au$ reactions (b=4.5 fm) at different beam 
energies as a function of time.}
\label{fig3}
\end{figure}
In Fig. 3 we show 
the evolution with time of the local temperature and the density for the 
spectator in semi-central $Au+Au$ collisions at various incident energies. It is seen that after the time when the spectator is 
fully developed  the densities and 
temperatures are rather independent of incident energy and remain rather constant for several tens of fm/c, making it an ideal system in 
order to study the thermodynamical evolution of low-density, finite temperature nuclear matter. In Fig. 4 we show evolution of pressure and density with time as a parameter.  We see that after about 45 fm/c the 
effective compressibility $K \sim \partial P/\partial \rho |_T$ becomes negative indicating that the system enters a region of spinodal instability and will subsequently break up into fragments. At this time we find 
densities of about $\rho \sim 0.4 - 0.5 \rho_0$ and $T \sim 5 - 6$ MeV, which is in good agreement with findings of the ALADIN group based on isotope 
thermometers \cite{aladin98,aladin99}. The figure also shows isotherms of the bulk matter EOS at temperatures in the range of the spectator temperature. It is seen that the finite system follows the bulk material rather closely.
\begin{figure}[ht]    
\unitlength1cm
\begin{center}
\begin{picture}(14,6)
\put(0.5,-1.0){\makebox{\epsfig{file=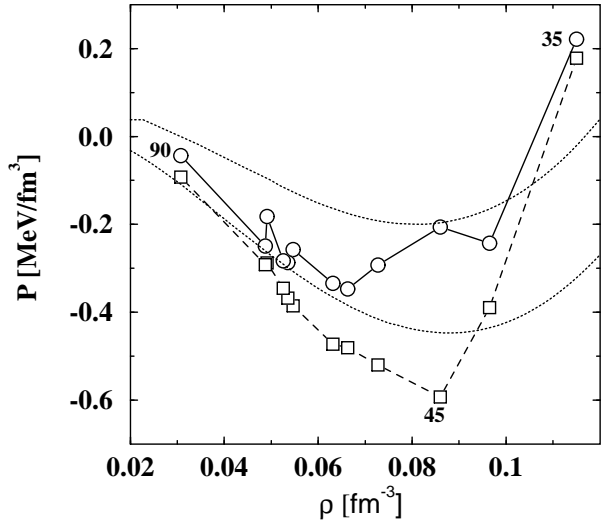,width=10.0cm}}}
\end{picture}
\end{center}
\caption{     
Density-pressure trajectory for the spectator matter in a semi-central 
Au on Au reaction at 600 A.MeV as in Fig.\,3. The dotted curves
are the nuclear matter equation of state for T = 5 and 9~MeV (lower and 
upper curve, respectively).
}
\end{figure}

Recently the ALADIN group has also determined kinetic energy spectra of 
spectator fragments \cite{aladin99} and has extracted slope temperatures using eq.(\ref{fit}). 
Corresponding spectra, generated in the coalescence model, were shown in Fig. 2 together with the blast model fits. 
 In Fig. 5 we show the slope 
temperatures separately for the different fragment masses and also the 
local temperature for comparison.  
 For the nucleons we have  $T_{slope} = (7.3 \pm 3.5)$ MeV which is close to the local temperature  $T_{loc}=(5 - 6)$ MeV (see Fig. 3), as one would expect.
On the other hand the slope temperatures of the fragments are considerably 
higher than those of the nucleons saturating for $A_f \ge 3$ 
around $T_{slope} \sim 17$ MeV. 
The experimental values from ALADIN \cite{aladin99} are also shown in Fig. 5. 
It can be seen that the slope temperatures from the theoretical calculations 
and those from the data agree extremely well.
\begin{figure}[ht]
\unitlength1cm
\begin{center}
\begin{picture}(14,5.5)
\put(1.5,-1.5){\makebox{\epsfig{file=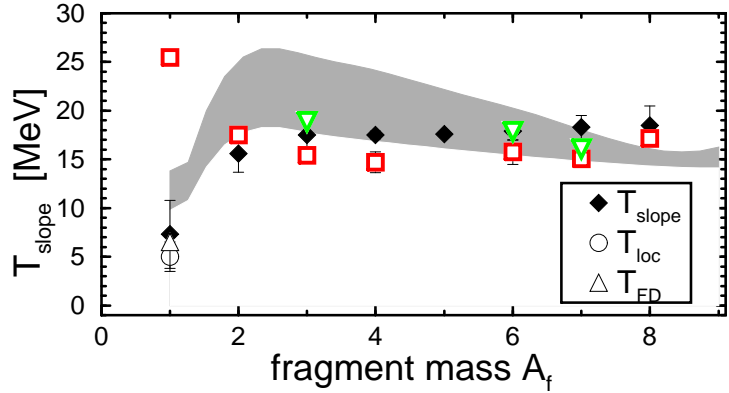,width=13.0cm}}}
\end{picture}
\end{center}
\caption{ Spectator slope temperatures for different fragment masses $A_{f}$ 
for the reaction as in Fig. 5 (diamonds). Also shown is the 
nucleon local temperature (circle) and the temperature obtained from a 
statistical model (gray band, see text). }
\label{fig5}
\end{figure}

We see that $T_{slope}$  for fragments  differs from that for nucleons and also from $T_{loc}$. Similar differences were seen by the ALADIN collaboration between   $T_{slope}$  from the fragment spectra and temperatures determined from isotope ratios  $T_{iso}$. 
The difference  has been interpreted in ref.\cite{aladin99} in terms 
of the Goldhaber model \cite{goldhaber,bauer}. When a system of fermions of given density and 
temperature breaks up the fragment momenta are approximately 
given by the sum of the momenta of the nucleons 
before the decay which leads to  energy distributions which resemble Maxwellians and thus contribute to the slope temperature. 
To check this argument in this case we therefore initialized statistically a system of the mass 
and temperature of the spectator, and subjected it to the 
same fragmentation procedure (coalescence) and to the same fit by eq. 
(\ref{fit}) as we did for the heavy ion collision. The slope
temperatures obtained from this 
statistical model are given in Fig. 5 as a band, which corresponds to 
initializations covering the range of values in Fig. 3. It 
is seen that the model qualitatively explains the increase in the slope 
temperature relative to the local temperature and the increase with fragment 
mass relative to that for nucleons. A similar conclusion was drawn in ref 
\cite{aladin99} using the results from ref. \cite{bauer}. 
This shows that $T_{slope}$  is {\it not} a thermodynamic temperature but simulates a  higher 
temperature. This effect has been called ''contribution of Fermi motion to 
the temperature''.

\subsection*{Participant matter}

The participant zone in a heavy ion collision constitutes another limiting, but still simple case for the investigation of the thermodynamical behaviour of 
nuclear matter. In contrast to the spectator zone one expects a 
compression-decompression cycle and thus richer phenomena with respect to 
fragmentation. The situation becomes particularly simple if we look at central 
collisions of symmetric systems \cite{eos95,reisdorf,fopi}.
We find that in $Au + Au$ at 600A MeV a very well developed radial flow pattern appears after 
about 20 fm/c  and the pressure becomes isotropic at about 35 fm/c indicating equilibriation.
The number of collisions drops to small values at about 40 fm/c. This condition we shall call (nucleon) freeze-out. Thus equilibration and freeze-out occur 
rather simultaneously. We find a density at this stage of about normal nuclear 
density and a (local) temperature of about $T_{loc} \sim 15$ MeV in the 
mid-plane of the reaction. 

We now also apply the blast model of eq. (\ref{fit}) to fragment spectra 
generated in 
the coalescence model at the end of the collision at about 90 fm/c. The results for the slope temperature $T_{slope}$ and the mean velocity $\beta_f$ are 
shown in Fig. 6 for a common fit to all fragments with $A_f \ge 2$ for different 
incident energies. These are compared to the corresponding values extracted by 
the EOS \cite{eos95} and FOPI 
\cite{reisdorf} collaborations. Within the 
uncertainities of the description by blast model fits, where the parameters  $T_{slope}$ and $\beta_f$ are not completely independent, there is qualitative 
agreement between calculation and experiment.
\begin{figure}[ht]
\unitlength1cm
\begin{center}
\begin{picture}(14,8)
\put(2.3,-0.5){\makebox{\epsfig{file=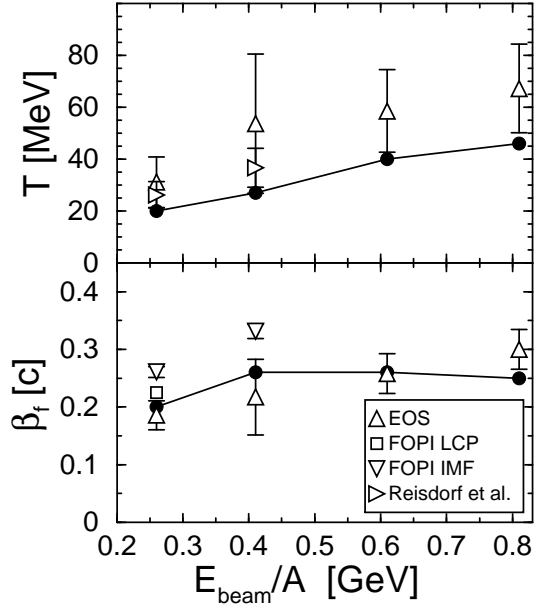,width=7.0cm}}}
\end{picture}
\end{center}
\caption{Slope temperatures (top) and radial flow velocities (bottom) 
from blast model 
fits to fragment ($A_f>1$) energy spectra 
for central collisions for different beam energies. 
They are compared to data from \protect\cite{eos95,reisdorf}.}
\label{fig6}
\end{figure}

As was done for the spectator we also apply the blast model separately for the
different fragment masses $A_f$. This is shown in Fig. 7 at 600 AMeV in the left 
column. We observe that slope temperatures increase and flow velocities fall with 
fragment mass in contrast to the behaviour for the spectator fragments in Fig. 5 
where $T_{slope}$ was about constant. A similar behaviour has been seen experimentally at 1 A.GeV in ref. \cite{eos95,reisdorf} 
and in calculations in ref. \cite{hombach98}.
Thus this behaviour cannot be interpreted as 
fragments originating from a common freeze-out configuration, i.e. from a 
fragmenting source. To arrive at an interpretation we have shown on 
the right column of Fig. 7 the local temperatures and flow velocities for 
different times 
before the nucleon freeze-out, i.e. for $t'=t_{freeze-out}-t$, with  
$t_{freeze-out} \sim 35 fm/c$. For fragment masses $A_f > 1$ the slope temperatures and velocities behave qualitatively very similar to the local temperatures and flow 
velocities at earlier times . 
This would suggest to interpret the results on the lhs of Fig. 7 
as signifying that 
heavier fragments originate at times earlier than the nucleon freeze-out. 
This may not be unreasonable
since in order to make a heavier fragment one needs higher densities 
which occur at earlier times and hence higher temperatures. 
However, this does not neccessarily imply that the fragments are really 
formed at this time, since fragments could hardly survive such high temperatures. But it could mean that these 
fragments carry information about this stage of the collision. Similar conclusions were drawn in calculations in the QMD model \cite{aichelin}. In any 
case it means that in the participant region fragments are {\it not} 
formed in a common equlibrated freeze-out configuration, and that in 
such a situation slope temperatures have to be interpreted with great caution.
\begin{figure}[ht]
\unitlength1cm
\begin{center}
\begin{picture}(14,8.3)
\put(0.5,-0.5){\makebox{\epsfig{file=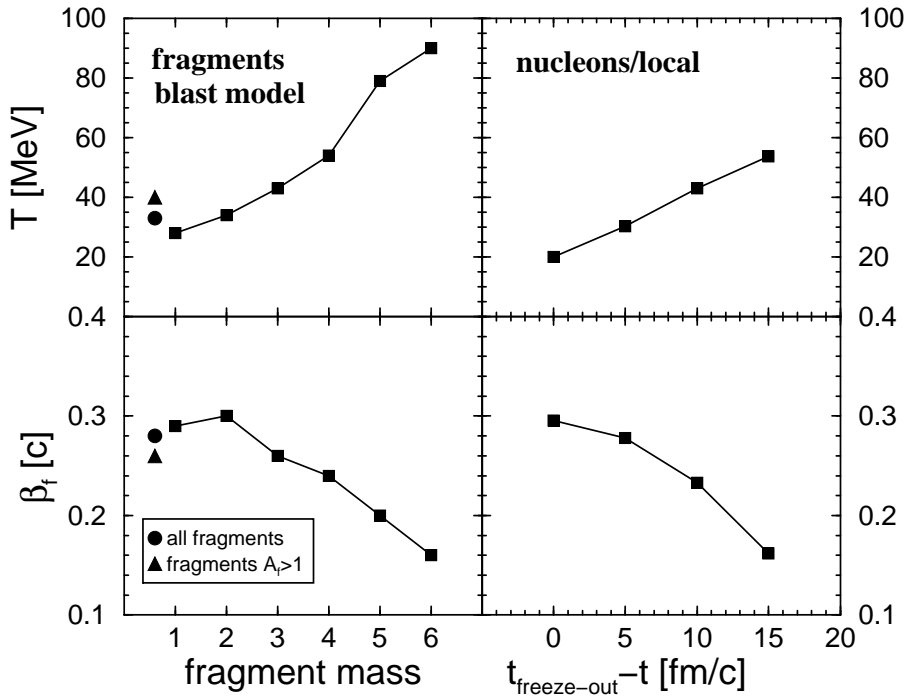,width=12.0cm}}}
\end{picture}
\end{center}
\caption{
Slope temperatures (upper row) and flow velocities (lower row) for the 
same reaction as in fig. 4 at $E_{beam}=0.6$ AGeV. In the left column 
 for blast model fits for different fragment masses and 
also for $A_f>1$ and for all fragments;
in the right column the local values from the momentum distribution 
at times before the freeze--out.}
\label{fig7}
\end{figure}

\subsection*{Summary}

To summarize fragmentation phenomena in heavy ion collisions are studied as a 
means to explore the phase diagram of hadronic matter. Here we have limited the discussion to the final stages of a collision and to the question of phase transitions and fragmentation, and  to determine the 
thermodynamical properties of the fragmenting source. One way to do this 
experimentally is to investigate fragment kinetic energy spectra. In theoretical 
simulations the thermodynamical state can be obtained locally in space and time from the phase space distribution. In this work we have compared this with the 
information obtained from the generated fragment spectra. We apply this method 
to the spectator and participant regions of relativistic $Au+Au$-collisions. 
We find that the spectator represents a well developed, equilibrated  and 
instable fragmenting source. The difference in temperature determined from 
the local momentum space (or experimentally from the isotope ratios) and from 
the kinetic energy spectra can be attributed to the Fermi motion in the 
fragmenting source as discussed in a Goldhaber model. 
In the participant region the local temperature at the 
nucleon freeze-out and the slope temperature from fragment spectra are 
different from those of the spectator. 
The slope temperatures rise with fragment mass which might indicate 
that the fragments are not formed in a common, equilibrated  source. \\ \\
We thank the ALADIN collaboration, in particular W. Trautmann and C. Schwarz, 
for helpful discussions. This work was supported in part by the German ministry of education and research BMBF under grant no. 06LM868I and grant no. 06TU887.

\end{document}